\documentclass{cimento}

\usepackage{graphicx}
\usepackage{graphics}

\usepackage{url,cite}
\title{Recent progress in QCD at the LHC}
\author{Juan Rojo~\from{ins:x}\from{ins:y}}
\instlist{\inst{ins:x}Physics Department, Theory Unit, CERN, CH-1211 Gen\`eve, Switzerland\\
\inst{ins:y}Rudolf Peierls Centre for Theoretical Physics, 1 Keble Road, University of Oxford, UK}

\PACSes{12.38.-t,12.38.Lg}

\begin{document}

\maketitle

\begin{abstract}
Perturbative Quantum Chromodynamics has experienced an impressive progress
in the last few years, boosted by the requirements of the LHC experimental
program.
In this contribution, I briefly review a selection of recent results
in QCD and LHC phenomenology, covering progress
in parton distribution functions, automation of NLO calculations, merging
and matching at NLO, new calculations at NNLO accuracy and their matching
to parton showers, and new developments
and techniques in jet physics and jet substructure tools.
\end{abstract}

\paragraph{Introduction}

With the discovery of the Higgs boson, high-energy physics has entered a completely
new era, where its main goals in the following years will be the detailed exploration
of the Higgs properties, such as its couplings and 
branching fractions, as well as the extensive exploration of the energy frontier
in the search for new physics beyond the Standard Model.
To optimize the scientific output of the LHC, 
a careful control on the theoretical uncertainties for the various
relevant signal and background processes is of paramount importance.
At an hadron collider such as the LHC, these uncertainties are dominated by 
strong interaction physics, perturbative Quantum Chromodynamics (QCD).
The progress of perturbative QCD in the last years, boosted by
the requirements of LHC data, has been impressive, and
it would be difficult to faithfully summarize it even in a  full review paper,
let alone in this short contribution.
Therefore, here I will limit myself to briefly present a  necessarily biased selection
of various important topics in perturbative QCD of particular
relevance for LHC physics, and  apologize
in advance for any omissions forced by the brevity of this contribution.

The structure of this contribution follows the flow diagram of a typical hadron collision.
First I will discuss progress in our understanding of the initial state
of hadron collisions, as encoded in the parton distribution functions
(PDFs) of the proton.
Then I will move to developments in NLO and NNLO calculational techniques
of perturbative matrix elements, and their consistent matching to
parton showers to achieve a realistic description of the 
final state.
Finally I will review progress in jet physics and jet substructure.

\paragraph{Parton Distribution Functions}
The initial state of hadronic collisions is the domain of the
parton distribution functions (PDFs) of the proton (see~\cite{Forte:2013wc,DeRoeck:2011na,Perez:2012um} for
recent reviews).
PDFs are an essential ingredient for LHC phenomenology: they limit the accuracy
with which it is possible to extract the Higgs boson couplings from data~\cite{Dittmaier:2011ti,Watt:2011kp},
degrade the reach of searches for massive new BSM particles at the 
TeV scale~\cite{AbelleiraFernandez:2012cc} and are the dominant 
systematic uncertainty in the determination of fundamental
parameters such as the $W$ boson mass, that are key ingredients for global consistency 
tests
of the Standard Model~\cite{Bozzi:2011ww}.
Being non-perturbative objects (although their scale dependence is
determined by the perturbative DGLAP evolution equations),
they need to be extracted from global fits to hard-scattering data.
Various PDF fitting collaborations provide regular updates
of their QCD analysis.
The most recent PDF sets from these collaboration are
{\sc ABM12}~\cite{Alekhin:2013nda}, {\sc CT10NNLO}~\cite{Gao:2013xoa}, 
{\sc HERAPDF1.5}~\cite{CooperSarkar:2011aa}, 
{\sc MSTW08}~\cite{Martin:2009iq} and {\sc NNPDF2.3}~\cite{Ball:2012cx}.
A recent benchmarking exercise, comparing the most updated PDF sets between them and with
LHC data, has been presented in~\cite{Ball:2012wy}. \\
\indent
A major recent development in PDF fits has been the inclusion of
a wide variety of LHC data.
While of the various PDF sets discussed above, only NNPDF2.3 and ABM12 already
include LHC data, other groups have performed studies of their impact
in tailored analysis.
LHC observables with PDF sensitivity 
range from traditional observables, like jet production~\cite{Watt:2013oha,Aad:2013lpa,Chatrchyan:2012bja} 
and inclusive electroweak boson production~\cite{Chatrchyan:2013mza,Aad:2012sb}, to processes that only
at the LHC have become available for PDF fits, like isolated photon
production~\cite{d'Enterria:2012yj}, $W$ production in association 
with charm quarks~\cite{Chatrchyan:2013uja,Aad:2014xca}, top quark pair 
production~\cite{Czakon:2013tha}, low and high mass
Drell-Yan production~\cite{CMSDY,Aad:2013iua} and $W$ and $Z$ production in association 
with jets~\cite{Malik:2013kba}, among others.
The data on $W$+$c$ production is particularly useful since it provides
a clean handle on the strange PDF $s(x,Q^2)$~\cite{Stirling:2012vh,Chatrchyan:2013uja,Alekhin:2014sya,Aad:2014xca}.
The use of cross-section ratios between different center-of-mass energies also provides
useful PDF sensitivity~\cite{Mangano:2012mh}, see for instance the ATLAS measurement of the ratio
of jet cross-sections between 7 and 2.76 TeV~\cite{Aad:2013lpa}.
While some of these PDF studies are being carried by 
the PDF fitting groups, recently also
the ATLAS and CMS collaborations themselves have developed an extensive program
of PDF determinations from their own measurements~\cite{Chatrchyan:2013mza,Aad:2013lpa,Aad:2012sb}.
This has been made possible by the development of
the open-source PDF fitting package {\sc HERAfitter}\footnote{\url{https://wiki-zeuthen.desy.de/HERAFitter}}, which has been used
in a variety of PDF-related analysis by ATLAS and CMS.\\
\indent
From the theory point of view, recent developments include
the combination of QED corrections together with the QCD ones for the DGLAP
evolution
in a PDF fit, {\sc NNPDF2.3QED} set~\cite{Ball:2013hta}, that includes also a determination of
 the photon PDF $\gamma(x,Q^2)$ from LHC data.
Such PDFs with QED effects are required by consistency for LHC calculations
when QED and electroweak effects are taken into account~\cite{Boughezal:2013cwa}.
The LO version of {\sc NNPDF2.3QED} has been used to produce an updates tune of the
{\sc Pythia8} Monte Carlo, the so-called Monash 2013 Tune~\cite{Skands:2014pea}.
The treatment of heavy quarks in global PDF fits has also been studied by various groups,
showing that the use of a fixed-flavor number scheme as compared to
a general-mass variable flavor
number one can explain most of the differences between the {\sc ABM} fits (based on the former)
and other PDF sets, that instead employ the latter~\cite{Ball:2013gsa,Thorne:2014toa}.
Also related to heavy quarks, thanks to the use of the running charm mass
$m_c(m_c)$ in DIS structure functions it is now possible~\cite{Gao:2013wwa,Alekhin:2012vu} 
to determine its value
from the HERA combined $F_2^c$ data with competitive uncertainties.
The important issue of the
sensitivity of the Higgs cross-section in gluon fusion with respect to
the choice of dataset has been investigated
by CT in Ref.~\cite{Dulat:2013kqa}, and by CT together with other groups in the upcoming
2013 Les Houches proceedings.
Finally, the possibility of an intrinsic charm component in the proton has been
recently revisited by the CT group~\cite{Dulat:2013hea}, and important constrains
here should be provided by LHC observables such as $Z$+$c$.

\paragraph{Automation of NLO calculations, Matching and Merging}
During many years, the needs for NLO calculations were summarized in the
so-called Les Houches wish-lists.
However, these have become rapidly obsolete with the progress in the automation
of NLO calculations by various groups, which makes the calculation of
NLO processes essentially a solved problem, and in the latest Les Houches
report it has been replaced with  NNLO and NLO electroweak wish-lists\footnote{The Les
Houches 2013 workshop proceedings, in preparation.}.
They key for the automation of NLO calculation has been on the one side, the development
of methods for subtraction of soft and collinear singularities in real emission
diagrams, such as in the {\sc MadFKS}~\cite{Frederix:2009yq} and {\sc Sherpa}~\cite{Gleisberg:2008ta} programs, 
and the corresponding progress in the computation of
virtual amplitudes, with tools such as {\sc GoSam}~\cite{Cullen:2011ac}, 
{\sc CutTools}~\cite{Ossola:2007ax}, {\sc MadLoop}~\cite{Hirschi:2011pa},
{\sc OpenLoops}~\cite{Cascioli:2011va}
and {\sc Helac-NLO}~\cite{Bevilacqua:2011xh}, just to name a few.
Despite this automation, and specially for high final state multiplicities,
tailored NLO calculations are still required for efficiency, such as those
provided by  {\sc BlackHat}~\cite{Bern:2013gka} and {\sc Njet}~\cite{Badger:2013yda}.
The state of the art of fixed-order NLO calculations is provided by the
recent computation of $pp\to W+5~{\rm jets}$ by the {\sc BlackHat}
collaboration~\cite{Bern:2013gka}.
Another related topic that has received attention recently is based around ideas
for improved NLO computations, such as
in the {\sc MINLO} approach~\cite{Hamilton:2012np}.
The basic idea here is
 defining an optimal central scale so that it a good choice all over the phase-space, 
and not only to compute the total rate or a single distribution.
The {\sc MINLO} approach
is specially suited for the matching of fixed-order calculations with
parton showers.

In parallel to the automation of NLO calculations, the matching of these
to parton showers has also been automated to a good extent.
As an illustration, the {\sc MadGraph5\_aMC@NLO} program~\cite{Frederix:2011zi} 
has recently been made
public, which essentially upgrades {\sc MadGraph5}~\cite{Alwall:2011uj} 
to the NLO level, including
the matching to various parton showers such as {\sc Pythia8}~\cite{Sjostrand:2007gs}.
Similar features are provided in other frameworks such as {\sc Powheg-Box}~\cite{Alioli:2010xd} and
{\sc Sherpa}~\cite{Gleisberg:2008ta}, which are widely used by the LHC experiments in the analysis of their data.

\begin{figure}[t]
\begin{center}
\includegraphics[width=0.80\textwidth]{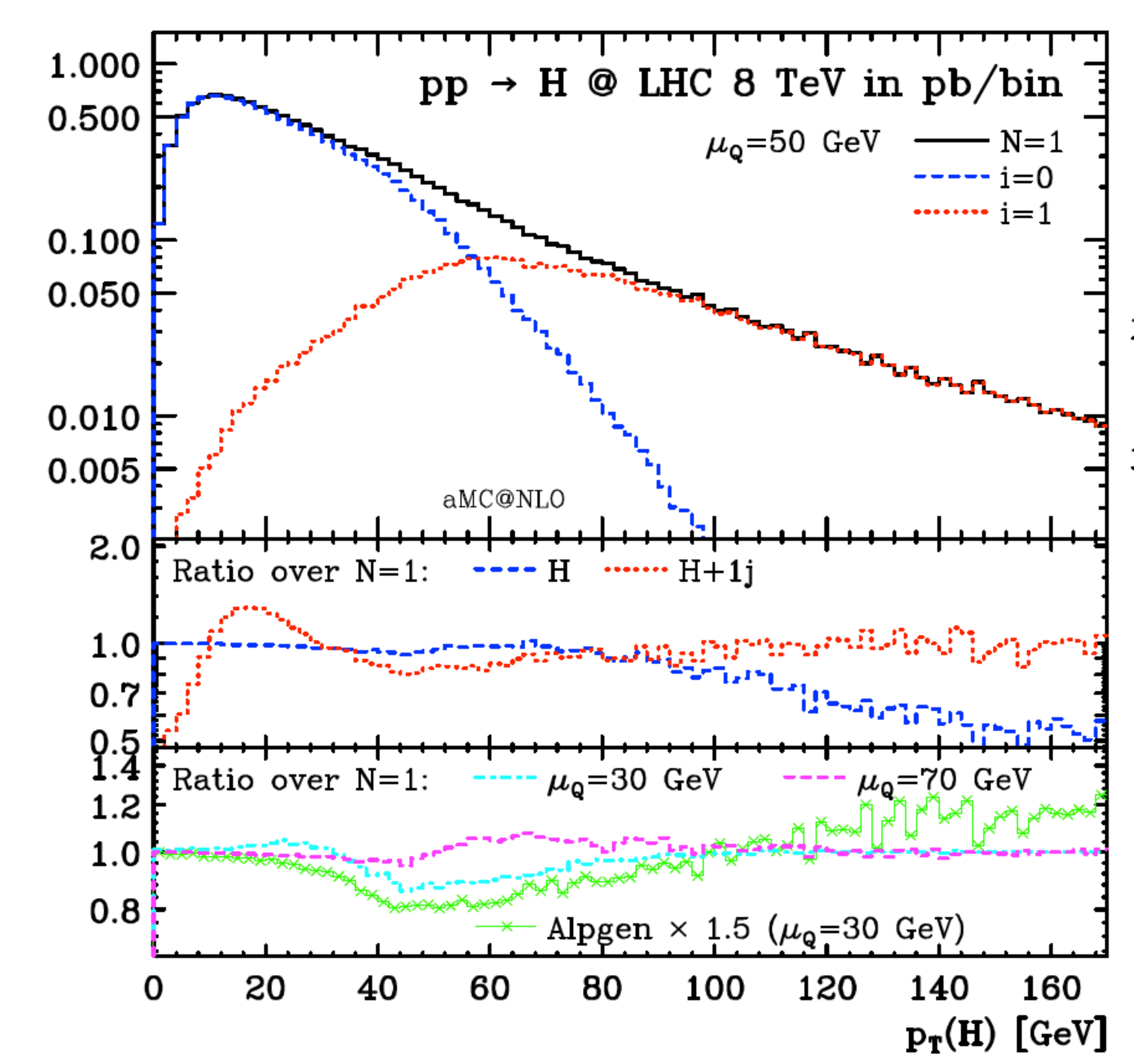}     
\end{center}
\caption{\small Illustration of the NLO merging procedure as
implemented in the {\sc FxFx} approach in Higgs plus jet production,
from Ref.~\cite{Frederix:2012ps} \label{fig:fxfx}}
\end{figure}

In order to achieve the best possible description of a realistic final state
of LHC collisions, it is required to match samples with different
parton level multiplicities.
While merging leading-order samples of different multiplicities (matched to parton showers) 
has been  well understood problems for more than a decade, with
various prescriptions available, such as {\sc CKKW}~\cite{Catani:2001cc}
and {\sc MLM}~\cite{Mangano:2001xp}, the extension of this merging procedure at NLO was much more challenging.
Recently, several approaches has been proposed for this combination of matching and merging,
including {\sc FxFx}~\cite{Frederix:2012ps}, {\sc MEPS@NLO}~\cite{Schonherr:2013cpa} 
and {\sc UNLOPS}~\cite{Lonnblad:2012ix}. These tools are important
in that they allow to carry out LHC phenomenology at the NLO level for all
exclusive processes of relevance.
As an illustration of the advantages of NLO merging, in Fig.~\ref{fig:fxfx} I show the results
of the NLO merging procedure in the {\sc FxFx} approach for Higgs production
in gluon fusion in association with one extra jet, from Ref.~\cite{Frederix:2012ps}.

At the level of precision that QCD calculations are achieving, it is becoming more
pressing to also include electroweak corrections to a variety of LHC processes.
Since purely weak corrections are enhanced by double logarithms
 of the type $\alpha_{W}\log ^2 s/M_W^2$, the
effect of these corrections becomes more important at the 
upcoming 13 TeV run of the LHC.
A comprehensive summary of the status of NLO weak calculations at hadron colliders
has been presented in Ref.~\cite{Mishra:2013una} in the context of the
Snowmass studies.
With a similar motivation, the possibility of including weak corrections in the parton shower~\cite{Christiansen:2014kba}, as well
as in the evolution of parton distributions~\cite{Ciafaloni:2001mu}, is also being studied.

\paragraph{The NNLO Revolution}
While NLO is the first order for which theory uncertainties in QCD calculations
are at the 10-20\% level, NNLO corrections are essential to get down to the percent
level and match the experimental accuracy of many LHC observables.
Until recently, only few QCD processes where available
at NNLO in a fully differential way, and these were restricted
to processes with either colorless initial or final state: for
hadron-hadron collisions, these included
Higgs production in gluon fusion, inclusive vector boson
production and di-photon production.

\begin{figure}[t]
\begin{center}
\includegraphics[width=0.54\textwidth]{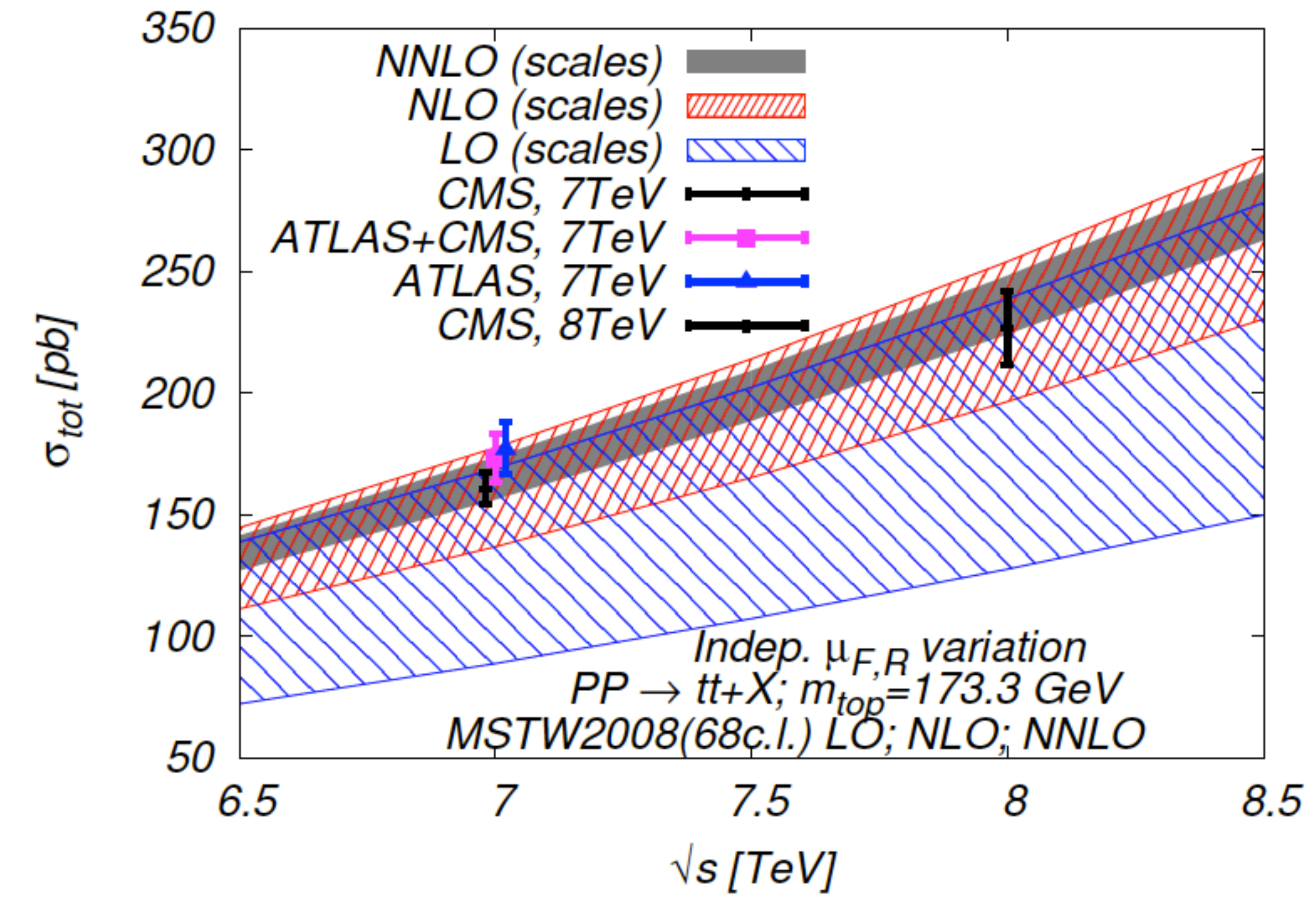}     
\includegraphics[width=0.42\textwidth]{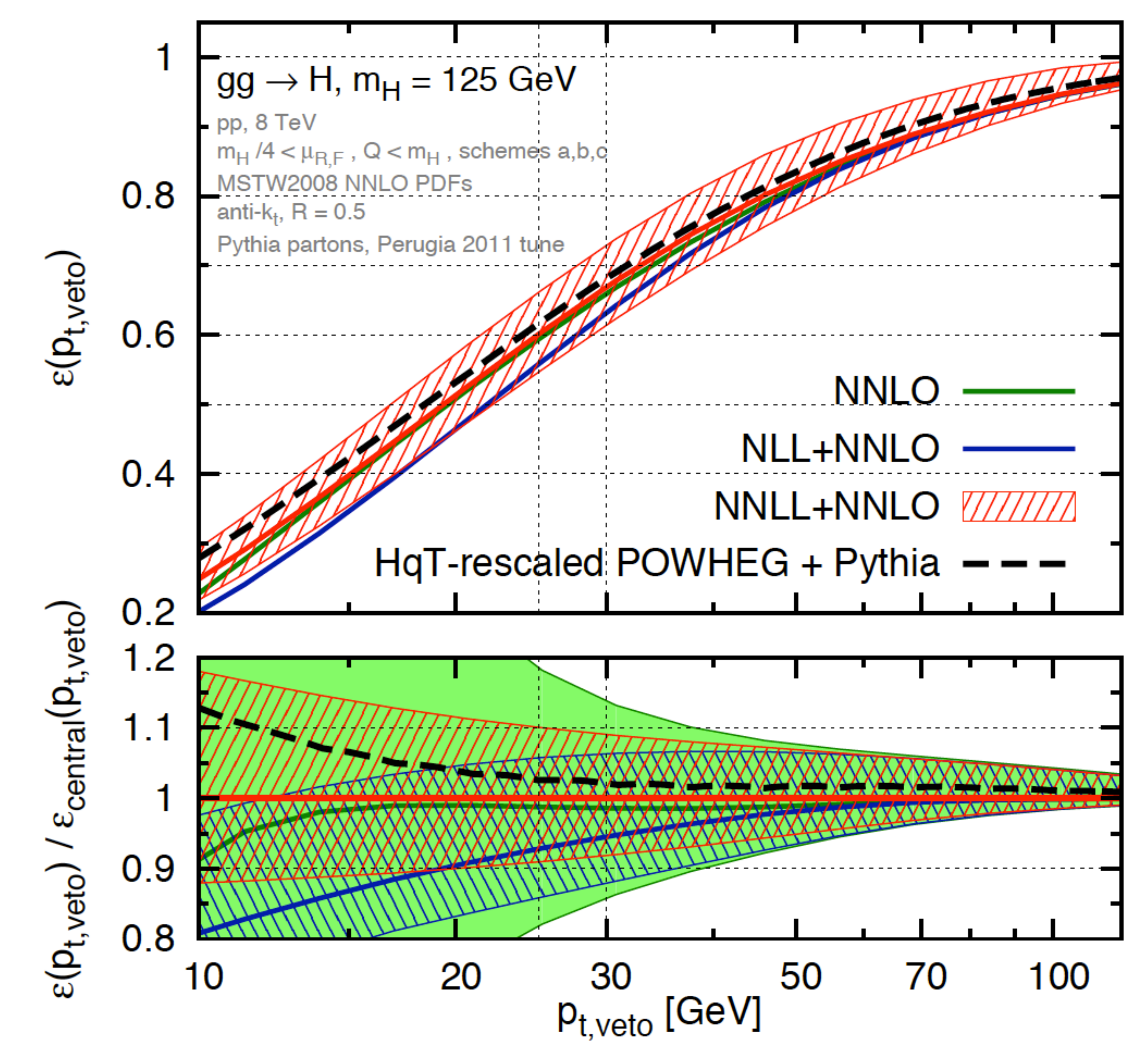}     
\end{center}
\caption{\small Left plot: the NNLO calculation of  $\sigma (t\bar{t})$ 
as a function of the collider center-of-mass energy.
Right plot: resummed calculation for the jet veto efficiency $gg\to H$, from~\cite{Banfi:2012yh}.
 \label{fig:nnlojet}}
\end{figure}

However, the development of new calculational techniques, such
as Antennae Subtraction~\cite{GehrmannDeRidder:2005cm} and Sector-Improved 
decomposition~\cite{Czakon:2010td,Boughezal:2011jf}, lead in 2013
to a breakthrough in NNLO QCD calculations,
and in particular for the first time it has become possible to
compute NNLO corrections for processes with both colored initial
and final states.
These landmark NNLO results include the fully differential
distributions for the
gluon-gluon initial state in dijet production~\cite{Currie:2013dwa} and 
for Higgs production in association with one jet~\cite{Boughezal:2013uia},
as well as the total cross-section for top quark pair
production~\cite{Czakon:2013goa}, where in this latter case 
 the differential distributions should follow soon. 
These results are an important milestone towards a new level of precision for
LHC phenomenology, for instance allowing to include consistently for the first time
jet and top pair production data into global NNLO fits, and using the
 NNLO $H+j$ calculation to reduce the theoretical
uncertainties in the determination of Higgs couplings.
As an illustration of the reduction in theory uncertainties in NNLO calculations,
in Fig.~\ref{fig:nnlojet}, taken from~\cite{Czakon:2013xaa},
 I show the predictions for the $t\bar{t}$ 
cross-section as a function of the collider center-of-mass energy.

Thanks to the improvement in calculational techniques discussed above,
 more NNLO results are expected in the near future.
At this point, the ultimate accuracy frontier for the QCD description
of LHC processes would be the matching of these NNLO calculations
with parton showers, in order to achieve the most precise theory description
of exclusive LHC final states.
This challenging problem has already seen encouraging progress
with different approaches aiming to establish NNLO+PS
as the next frontier of QCD calculations.
Two of these approaches are the one of the {\sc Geneva} group, based
on soft-collinear effective theory~\cite{Alioli:2013hqa}, and the other is based on generalizing the
{\sc MINLO} approach to higher orders~\cite{Hamilton:2013fea}.

\paragraph{Jet Physics and Jet Substructure}

Jets are ubiquitous in hadronic collisions~\cite{Salam:2009jx}, an essential tool for virtually all
LHC analysis, from SM measurements and Higgs physics to BSM searches.
The current paradigm for jet reconstruction at the LHC is based on the
use of the anti-$k_T$
algorithm~\cite{Cacciari:2008gp} with jet radius $R$ in a range between 0.4 and 0.7.
Virtually all relevant jet algorithms and jet reconstruction tools,
from the most basic to the more advanced, are available through the
{\sc FastJet} framework~\cite{Cacciari:2011ma}, either part of the core code or as part of the
{\sc FastJet contrib} project\footnote{\url{http://fastjet.hepforge.org/contrib/}}.

Recently, substantial effort has been devoted to the understanding
of the theoretical uncertainties that arise in 
perturbative calculations which involve the presence of jet vetoes.
This is particularly important for Higgs analysis where events are classified in
bins of different jet multiplicity, with the motivation to disentangle
the different production channels, such as gluon-fusion from
vector-boson-fusion.
In this context, various resummed calculations have been proposed~\cite{Banfi:2012yh,Becher:2013xia,Boughezal:2013oha}, 
which allow to reduce perturbative uncertainties in the jet veto
efficiencies, both for the $H$+0 jet and the $H$+1 jet processes.
As an illustration, in Fig.~\ref{fig:nnlojet} I show the results for the
resummed calculation for the jet veto efficiency in $gg\to H$, taken from Ref.~\cite{Banfi:2012yh}.

Another area that has witnessed intense activity recently is that of jet substructure.
In the decays of boosted resonances, either SM particles like $W$ or top
quarks~\cite{Plehn:2009rk}, the Higgs boson~\cite{Butterworth:2008iy}, or for heavy new particles present in many BSM
scenarios, the
 decay prongs are often
collimated into a single jet, and jet substructure techniques can
be used to discriminate these {\it fat} jets with respect to the 
standard QCD jets.
With this motivation, a variety of jet substructure tools has been proposed
to sharpen interesting signals and at the same time tame the overwhelming QCD backgrounds.
Detailed reviews of the developments in this area can be found in the proceedings
if the BOOST series of workshops~\cite{Abdesselam:2010pt,Altheimer:2013yza}.
A related interesting issue is the matching of the resolved and boosted regime into
a unified analysis strategy, see Ref.~\cite{Gouzevitch:2013qca} 
for a proposal to address this problem.

However, I should emphasize that it is crucial to avoid using  blindly these tools.
Indeed,  it
is essential to back them not only with Monte Carlo studies but also
with analytical calculations, such as the recent studies of Ref.~\cite{Dasgupta:2013ihk}.
Thanks to these analytical insights, further-improved jet taggers and groomers
can be obtained, and their results trusted with higher confidence.
In addition, 
in parallel with theoretical and computational developments, the validation of the various substructure
tools on real LHC data is an essential prerequisite before they can be safely applied
to searches of new physics in boosted final states.

\paragraph{Outlook}

It should be clear from the above discussion that the recent 
progress in perturbative QCD
has been impressive, and that even better results are being prepared for the years to come.
These include parton distributions based only on collider data and including NNLO QCD and NLO electroweak
corrections, the maturity of the NLO merging approaches, a wider range of NNLO calculations
and their matching to parton showers, and the quantitative improvement of jet substructure taggers
and groomers, among many others.
All these developments will lead to a new generation of
precision for our study of LHC
processes and the exploration of the laws of Nature at the energy frontier.

\providecommand{\href}[2]{#2}\begingroup\raggedright\endgroup

\end{document}